\documentclass[dvips,epsfig]{article}
\usepackage[T1]{fontenc}
\input epsf.tex
\makeatletter

\providecommand{\LyX}{L\kern-.1667em\lower.25em\hbox{Y}\kern-.125emX\@}

\newcommand{\lyxaddress}[1]{
  \par {\raggedright #1 
  \vspace{1.4em}
  \noindent\par}
}

\usepackage[T1]{fontenc}

\makeatletter

\usepackage[T1]{fontenc}

\makeatother

\begin{document}

\title{\hfill{}PM/02-24\\
 Supersymmetric penguin contributions to the process \protect\protect\protect\( B_{d}\longrightarrow \Phi K_{s}\protect \protect \protect \)
in SUSY GUT with right-handed neutrino.}

\author{M.B.Causse\thanks{
Marie-Bernadette.Causse@gamum2.in2p3.fr 
}}

\maketitle

\lyxaddress{Groupe d'Astroparticules de Montpellier Université Montpellier II Place Eugene
Bataillon, 34095 Montpellier Cedex 5}

\begin{abstract}
The \( B_{d}\longrightarrow \Phi K_{s} \) process is without tree level decay
amplitude . Then the SUSY contributions may be significant . We calculate the
neutralino and gluino penguin contributions to the CP violating phase in \( SU(5) \) grand unified theory with right-handed neutrino \( SU(5)_{RN} \)
. We find that the SUSY contributions change the phase in \( B_{d}\longrightarrow \Phi K_{s} \). 
\end{abstract}

\section{Introduction}

In the minimal Standard Model(SM) lepton flavor is conserved . Then Lepton Flavor
Violation (LFV) and flavor changing neutral current (FCNC) is forbidden at tree
level . Therefore, the flavor mixing is controlled by the Cabibbo Kobayashi
Maskawa (CKM) matrix element . So the phase in the CKM matrix is the only source
of the CP violation in the Standard Model (neglecting the strong CP problem)
.

However, in supersymmetric (SUSY) models there is a new source of flavor mixing
in the mass matrices of SUSY partners for leptons and quarks . In the context
of minimal Supergravity (mSUGRA), where SUSY breaking parameters are assumed
to be flavor blind at the Planck scale, the flavor mixing is controlled by the
CKM matrix elements . As a result, the maximal deviation from the SM in the
CP violating parameters is of the order of \( 10\% \), while the SUSY contribution
to \( b\longrightarrow s\gamma  \) process can be much more important\cite{1}
. In the GUT scenario, there are additional contributions to FCNC/LFV process
from GUT interaction\cite{2} . As for LFV processe the \( \mu \longrightarrow e\gamma  \)
branching ratio is closed to the current experimental bound, especially for
\( SO(10) \) model \cite{3} . Experimental evidences of neutrino oscillation
indicate the existence of small neutrino mass and large flavor mixing in the
lepton sector . A natural explanation for small neutrino masses is the seesaw
mechanism\cite{4} . In this mechanism, heavy right-handed neutrinos are introduced
. These neutrinos have Majorana mass term and new Yukawa interactions . In SUSY
models with a neutrino mass generation mechanism of seesaw type, Yukawa coupling
constants among the Higgs doublet, lepton doublets and right-handed neutrinos
could induce large flavor mixing effects in the slepton sector . In the Supersymmetric
Grand Unified theory (GUT) with right-handed neutrino, above the GUT scale,
the right-handed down type squarks couple to the right-handed neutrinos . Then,
due to renormalization group effect, flavor violation in the slepton sector
may be transferred to the right-handed down type squark mass matrix\cite{5},
which affects the CP violation in the \( B \) decay .\\
 The \( B_{d}\longrightarrow \Phi K_{s} \) process is without tree level decay
amplitude . Then the SUSY contributions may be significant . The SUSY contributions, to the CP violation 
in this process, are gluino, neutralino and chargino . In the \( B_{d}\longrightarrow \Phi K_{s} \) process, gluino and neutralino 
interact with down type squark-quark (fig 10,11 ), whereas the charginos interact with up type squarks . So in our model, \( SU(5)_{RN} \), 
only gluino and neutralino contributions are concerned by flavor violation in the slepton sector .  

In this context we calculate the neutralino and gluino penguin contributions
to the CP violation in the process \( B_{d}\longrightarrow \Phi K_{S} \) . Then, the SUSY contributions
can be large and the phase in the decay amplitude can be as large as \( O(0.1) \)
which is much larger than the SM . In our calculation we apply the multimass
insertion method \cite{6} . We take into account the mixing of the squark eigenstates
as well as the mixing in the neutralino sector and we consider the case of large
mixing between second and third generation, suggested by the atmospheric neutrino
problem \cite{10} . The calculation of the neutralino contribution is new .
Previous gluino calculation can be found in the reference \cite{9} equation (A.16), where mass insertion method has been used\footnote{%
In this method an average squark mass is taken into account only.
}
 . In this calculation to obtain the results presented in figure (2) by the author, we must take
the derivative of the equation (A.16) of the reference \cite{9} . Consequently,
we have also calculated the gluino contribution, with multimass insertion method ( 
where the left and right squark masses are taken into account) . 

This paper is organized as follows . In section 2 the \( SU(5) \) grand unified
theory with right-handed neutrino \( SU(5)_{RN} \) is introduced . In section 3,
we give the explicit expressions of the amplitudes associated to the neutralino
and gluino contributions to the decay \( B_{d}\longrightarrow \Phi K_{s} \)
(\( b\longrightarrow s\Phi  \)) . Finally, in section 4 we find the numerical
results on the SUSY contributions to the decay phase for different values on
SUSY parameters . In section 5 we conclude . In annex, the analytic expressions
for the Feynman integrals, which appear in the evaluation of the amplitudes,
are given .

\section{\protect\protect\protect\( SU(5)\protect \protect \protect \) grand unified
theory with right-handed
\newline 
neutrinos \protect\protect\protect\( SU(5)_{RN}\protect \protect \protect \)}

We consider the supersymmetric \( SU(5) \)grand unified theory (GUT) with right-handed
neutrinos, as an extension of the Minimal Supersymmetric Standard Model with
right-handed neutrinos (MSSMRN)\cite{5}, \cite{6}, \cite{7} . Above, the
GUT scale the superpotential is :

\begin{equation}
\label{pot1}
\begin{array}{c}
W=\frac{1}{8}\Psi _{i}\, [Y_{U}]_{ij}\, \Psi _{j}\, H\, +\, \Psi _{i}\, [Y_{D}]_{ij}\, \Phi _{j}\, \overline{H}\, \\
+\, N_{i}\, [Y_{N}]_{ij}\, \Phi _{j}\, H\, +\, \frac{1}{2}N_{i}\, [M_{N}]_{ij}\, N_{j}
\end{array}
\end{equation}

where \( \Psi _{i} \) (the quark doublet), \( \Phi _{i} \) (the charged lepton
doublet) and \( N_{i} \) (the right-handed neutrino) are 10, \( \overline{5} \)
and singlet matter fields of \( SU(5) \) in i-th generation, while \( \overline{H} \)
and H are Higgs fields in \( 5^{*} \) and 5 representations, such that : \( H=(H_{2,}H_{c}) \)
and \( \overline{H}=(H_{1},\overline{H}_{c}) \) . Finally, \( H_{1}\, ,\, H_{2} \)
are the Higgs multiplets of the MSSM and \( H_{c}\, ,\, \overline{H}_{c} \)
are two different colored Higgs multiplets . Here \( M_{N} \) is the Majorana
mass matrix for right-handed neutrinos . For simplicity, we chose degenerate
Majorana mass matrix for the right-handed neutrinos :

\begin{equation}
\label{eq1}
[M_{n}]_{ij}=M_{\nu _{R}}\, \delta _{ij}
\end{equation}
 The Yukawa matrix \( Y_{U} \) is a complex symmetric matrix, while \( Y_{D} \)
and \( Y_{N} \) are complex . It is convenient to work in the basis where \( Y_{D} \)
is diagonal . Then we have :

\[
Y_{U}=V_{Q}^{\dagger }\, \widehat{\Theta }_{Q}\, \widehat{Y}_{U}\, V_{Q}\: ,\: Y_{D}=\widehat{Y}_{D}\: and\: Y_{N}=\widehat{Y}_{N}\, V_{L}\, \widehat{\Theta }_{L},\]

\( \widehat{Y}_{U}\, ,\, \widehat{Y}_{D}\, and\, \widehat{Y}_{N} \) are real
diagonal matrices:

\[
\widehat{Y}_{U}=diag(y_{u},y_{c},y_{t})\, ,\, \widehat{Y}_{D}=diag(y_{d},y_{s},y_{b})\, ,\, \widehat{Y}_{N}=diag(y_{\nu 1},y_{\nu 2},y_{\nu 3}).\]
 The \( \widehat{\Theta }'s \) are diagonal phase matrices :

\( \left\{ \begin{array}{c}
\widehat{\Theta }_{Q}=diag[\exp (i\phi _{1}^{(Q)}),\exp (i\phi _{2}^{(Q)}),\exp (i\phi _{3}^{(Q)})]\\
\\
\widehat{\Theta }_{L}=diag[\exp (i\phi _{1}^{(L)}),\exp (i\phi _{2}^{(L)}),\exp (i\phi _{3}^{(L)}]
\end{array}\right.  \)

such as: 
\[
\phi _{1}^{(Q)}+\phi _{2}^{(Q)}+\phi _{3}^{(Q)}\, =\, \phi _{1}^{(L)}+\phi _{2}^{(L)}+\phi _{3}^{(L)}\, =\, 0.\]
 Moreover, \( V_{Q} \) and \( V_{L} \) are unitary mixing matrices parametrized
by three mixing angles and on CP violating phase .

At GUT scale \( (M_{GUT}\approx 1.9\times 10^{16}GeV) \), the GUT gauge symmetry
is spontaneously broken into the SM one : \( SU(5)_{GUT}\rightarrow SU(3)_{c}\times SU(2)_{L}\times U(1)_{Y} \)
. So, at low energy, \( \Psi ^{i} \) contains \( Q^{i}(3,2,\frac{1}{6}) \),
\( \overline{U}^{i}(\overline{3},1,-\frac{2}{3}) \) and \( \overline{E}^{i}=(1,1,1) \),
quark and lepton superfields . \( \overline{\Phi }^{i} \) includes \( \overline{D}^{i}(3,1,\frac{1}{3}) \)
and \( L^{i}=(1,2,-\frac{1}{2}) \) quark and lepton superfields . \( \overline{N}^{i} \)
is a singlet of \( SU(3)_{c}\times SU(2)_{L}\times U(1)_{Y} \) . The representations
for SU(3), SU(2) groups and \( U(1)_{Y} \) charge are indicated in parenthesis
. Then the GUT multiplets in terms of MSSM field are:

\begin{equation}
\label{eq2}
\Psi _{i}\simeq \{Q,V_{Q}^{\dagger }\widehat{\Theta }_{Q}\overline{U},\widehat{\Theta }_{L}\overline{E}\}_{i}\, ,\, \Phi _{i}\simeq \{\overline{D},\widehat{\Theta }^{\dagger }_{L}L\}_{i}.
\end{equation}
 The superpotential(\ref{pot1}) becomes:

\begin{equation}
\label{pot2}
\begin{array}{c}
W=Q_{i}[V_{Q}^{\dagger }\widehat{Y}_{U}]_{ij}\overline{U}_{j}H_{2}\, +\, Q_{i}[\widehat{Y}_{D}]_{ij}\overline{D}_{j}H_{1}\, +\, \overline{E}_{i}[\widehat{Y}_{E}]_{ij}L_{j}H_{1}\\
+N_{i}[\widehat{Y}_{N}V_{L}]_{ij}L_{j}H_{2}\, +\, \frac{1}{2}M_{\nu _{R}}N_{i}N_{i}
\end{array}.
\end{equation}
 In the \( SU(5) \) limit, the naive GUT relation is predicted \( \widehat{Y}_{E}=\widehat{Y}_{D} \)
at the GUT scale . Although this relation gives a reasonable agreement for \( m_{b} \)
and \( m_{\tau } \), it is well know that the mass ratio of down-type quarks
and charged leptons in the first and the second generation can not be explained
in this way\footnote{%
One possibility to remedy this defect is to introduce higher dimensional operators
suppressed by the Planck scale. 
} . Then, in (\ref{pot2}), the unitary matrix \( V_{Q} \) becomes the CKM matrix
:\( V_{Q}\simeq V_{CKM} \) . The left-handed neutrino mass matrix induced by
seesaw mechanism is :

\begin{equation}
\label{eq3}
[m_{\nu }]_{ij}\, =\frac{v^{2}\sin ^{2}(\beta )}{2M_{\nu _{R}}}[Y_{N}]_{ij}^{2}\, =\frac{v^{2}\sin ^{2}(\beta )}{2M_{\nu _{R}}}[V^{\dagger }_{L}\widehat{Y}^{2}_{N}V_{L}]_{ij}
\end{equation}
 with \( v^{2}=2(<h_{1}>^{2}+<h_{2}>^{2})\approx (246Gev)^{2} \) and \( \tan (\beta )=\frac{<h_{2}>}{<h_{1}>} \)
. Then \( V_{L} \) plays the role of the neutrino mixing matrix and the neutrino
Yukawa coupling is :

\begin{equation}
\label{eq4}
y_{\nu }=M^{1/2}_{\nu _{R}}\times f(V_{L},m_{\nu },\tan (\beta )).
\end{equation}
 For a large \( M_{\nu _{R}} \) the equation(\ref{eq4}) becomes non-perturbative
below the Planck scale . In terms of MSSM fields, the third term in the superpotential(\ref{pot1})
can be written as :

\begin{equation}
\label{pot3}
W=N_{i}[\widehat{Y}_{N}V_{L}\widehat{\Theta }_{L}]_{ij}\overline{D}_{j}H_{c}+.....=\sum _{i,j}\, y_{\nu _{i}}[V_{L}]_{ij}\exp (i\phi _{j}^{(L)})N_{i}\overline{D}_{j}H_{c}\, +\, .....
\end{equation}
 The phases \( \phi ^{(L)} \) and \( \phi ^{(Q)} \) remain in the colored
vertices . The effect of the colored Higgs is negligible at tree level, but
affects the soft SUSY breaking mass parameters through the RG effect and generates
off-diagonal elements in the mass matrix of down-type squarks . In this model
we assume the universality of the scalar mass (\( m_{0} \)) at the reduced
Planck scale \( M_{*}\simeq 2.4\times 10^{18}Gev \) . Thus, we adopt the model
so-called mSUGRA . Due to the neutrino Yukawa matrix, non-vanishing off diagonal
elements are generated . The off-diagonal elements in the mass matrix of right-handed
down-type squarks (\( \widetilde{D} \)) are approximately :

\begin{equation}
\label{eq5}
[m^{2}_{\widetilde{d}_{R}}]_{ij}\simeq -\frac{1}{8\pi ^{2}}\exp [-i(\phi ^{(L)}_{i}-\phi _{j}^{(L)})]\sum _{k}y^{2}_{\nu _{k}}[V_{L}^{*}]_{ki}[V_{L}]_{kj}\times (3m^{2}_{0}+a^{2}_{0})\log (\frac{M_{*}}{M_{GUT}})
\end{equation}
 where \( a_{0} \) is the universal A parameter and R stands for right component
. For the left-handed down-type squarks, the off-diagonal elements in the mass
matrix are generated by the top Yukawa coupling, such as:

\begin{equation}
\label{eq6}
[m^{2}_{\widetilde{d}_{L}}]_{ij}\simeq -\frac{1}{8\pi ^{2}}y^{2}_{t}[V_{CKM}^{*}]_{ti}[V_{CKM}]_{tj}\times (3m^{2}_{0}+a^{2}_{0})\times (3\log (\frac{M_{*}}{M_{GUT}})+\log (\frac{M_{GUT}}{M_{weak}}))
\end{equation}
 where L stands for left component . These off-diagonal elements mass matrix
\( [m^{2}_{\widetilde{d}_{R}}]_{i3} \) and \( [m^{2}_{\widetilde{d}_{L}}]_{i3} \)
are coefficients of \( \Delta B\neq 0 \) operators, they change the standard
model predictions to the mixing and the decay of the B-mesons .

In this context we calculate the neutralino and gluino penguin contributions
to the CP violation in the process : \( B_{d}\longrightarrow \Phi K_{s}. \)

\section{Neutralino and gluino penguin contributions to
\newline
\protect\protect\protect\( B_{d}\longrightarrow \Phi K_{s}\protect \protect \protect \)}

The operators contributing to \( B_{d}\longrightarrow \Phi K_{s} \) \cite{8}
have a structure like \( (\overline{s}b)(\overline{s}s) \), then they are induced
at the one-loop level in the standard model . Therefore the SUSY contributions
may be significant . The SUSY contribution to the decay amplitude for \( B_{d}\longrightarrow \Phi K_{s} \)
comes from the penguin and box diagrams . In our analysis, we only take into
account the chromo-dipole contribution because, as is mentioned in reference\cite{9},
box and color-charge form factor contributions have the same size but the opposite
sign\footnote{%
In the neutralino case, the wino and higgsino couplings have the same sign as gluino ones . 
Then neutralino and gluino contributions have the same behaviour . For chargino it is different .
}
. Consequently, the \( \Delta B=1 \) effective Lagrangian considered takes
the following form:

\begin{equation}
\label{L1}
L_{eff}=m_{b}C^{DM}_{R}T_{ab}^{A}\overline{s}^{a}[\gamma ^{\mu },\gamma ^{\nu }]P_{L}b^{b}G_{\mu \nu }^{a}+(L\leftrightarrow R)+h.c
\end{equation}
 where \( m_{b} \) is the bottom quark mass,\( T_{ab}^{A} \) is the \( SU(3)_{c} \)
generator, \( G_{\mu \nu }^{a} \) is the gluon field and a and b the color
indices.

We have calculated the neutralino and gluino contributions to the coefficient \( C^{DM}_{R} \) . We apply the multimass insertion method .The calculation of the neutralino
contribution is new . Previous gluino calculation can be found in the reference \cite{9} equation(A.16), where mass insertion method has been 
used . The results are very large and to obtain the results presented in figure(2) by the author, we must take
the derivative of the equation(A.16) of the reference \cite{9} . Consequently,
we have also calculated the gluino contribution, with multimass insertion method, to the chromo-dipole operator .

\subsection{Neutralino contribution}

We assume neutrino and quark mass hierarchy and \( y_{\nu _{3}}\gg y_{\nu _{2}}\gg y_{\nu _{1}} \)
. From the atmospheric neutrino observation we take :

-for the neutrino tau mass :\( m_{\nu _{\tau }}=0.07ev \)

-and large mixing angle :\( V_{L32}=-\frac{1}{\sqrt{2}}. \)

Thus, from the equations (\ref{eq5}) and (\ref{eq6}) we have :

\begin{equation}
\label{eq7}
[m_{\widetilde{d}_{R}}^{2}]_{32}\approx -\frac{1}{8\pi ^{2}}\exp (-i(\phi ^{(L)}_{3}-\phi _{2}^{(L)}))y^{2}_{\nu _{3}}[V_{L}^{*}]_{33}[V_{L}]_{32}\times (3m^{2}_{0}+a^{2}_{0})\log (\frac{M_{*}}{M_{GUT}})
\end{equation}
 and 
\begin{equation}
\label{eq8}
[m^{2}_{\widetilde{d}_{L}}]_{32}\simeq -\frac{1}{8\pi ^{2}}y^{2}_{t}[V_{CKM}^{*}]_{tb}[V_{CKM}]_{ts}\times (3m^{2}_{0}+a^{2}_{0})\times (3\log (\frac{M_{*}}{M_{GUT}})+\log (\frac{M_{GUT}}{M_{weak}})).
\end{equation}
 These two terms contribute to the decay amplitude:

\begin{equation}
\label{eq9}
M_{B_{d}\rightarrow \phi K^{0}}^{(SUSY)}=<\phi K^{0}|L_{eff}|\overline{B}_{d}>.
\end{equation}
 The right-right non-diagonal term(eq.\ref{eq7}) depends on the neutrino phase
and is dominant compare to the left-left one (eq.\ref{eq8}) because \( \left| [V_{CKM}]_{ts}\right| \ll \left| [V_{L}]_{32}\right|  \)
. So, we only consider the right-right contribution to the \( M_{B_{d}\rightarrow \phi K^{0}}^{(SUSY)} \)
decay amplitude . Feynman rules used for the calculation are summarized in figures(10)-(11)
. In annex, we have the Feynman integrals . In the following we note:\( (\overline{m}^{2})_{32}=[m_{\widetilde{d}_{R}}^{2}]_{32} \)
.The three kind of diagrams contributing to the decay are illustrated in figure 1 and figure 2 .

From the diagrams illustrated in figure1 we obtain :

+for the diagrams 1(a) and 1(b)

\begin{equation}
\label{eq10}
\begin{array}{c}
\begin{array}{c}
T_{1a-b}=\frac{\sqrt{2}g^{2}_{2}g_{3}\tan (\theta _{w})\times (m_{b}+m_{s})M_{1\, }\mu \, \overline{(m}^{2})_{32}}{3\times 16\pi ^{2}[\overline{m}_{\widetilde{b}R}^{2}-\overline{m}_{\widetilde{s}R}^{2}][M_{1}^{2}-\mu ^{2}]}\\

\end{array}\\
\\
\times \sum ^{1}_{n=0}\frac{(-1)^{n}}{\overline{M}^{2}_{n}}\left[ \frac{FC_{1}(X_{\widetilde{w}_{n}})+FC_{1}(X_{\widetilde{\mu }_{n}})}{\overline{M}_{n}^{2}}+\frac{FN_{1}(X_{\widetilde{w}_{n}})-FN_{1}(X_{\widetilde{\mu }_{n}})}{M_{1}^{2}-\mu ^{2}}\right] \\

\end{array}
\end{equation}

with \( \overline{M}_{n}^{2}=\left\{ \begin{array}{c}
\, \, \overline{m}_{\widetilde{b}_{R}}^{2}\, for\, n=0\\
\overline{m}_{\widetilde{s}_{R}}^{2}\, for\, n=1
\end{array}\right.  \)and \( X_{\widetilde{w}_{n}}=\frac{M_{1}^{2}}{\overline{M}_{n}^{2}} \) and
\( X_{\widetilde{\mu }}=\frac{\mu ^{2}}{\overline{M}_{n}^{2}} \), where \( M_{1} \),
\( \mu  \) are the bino and higgsino masses, \( \overline{m}_{\widetilde{b}_{R}}^{2} \)
and \( \overline{m}_{\widetilde{s}_{R}}^{2} \) stand for the diagonal components
of the bottom and strange squark mass matrix (\( m_{s} \) is the strange quark
mass) . The coupling expressions are :

\( g_{2}^{2}=\frac{\alpha _{em}(M_{Z})}{4\pi \sin ^{2}(\theta _{w})}\, ,\, g_{3}^{2}=\frac{\alpha _{s}(M_{Z})}{4\pi } \),
q is the outgoing gluon momentum and p the bottom quark one .

+on the diagrams 1(c)-1(d), there is a flip of chirality on the bottom leg then,

\begin{equation}
\label{eq11}
T_{1c-d}=-\frac{g^{2}_{2}g_{3}\, m_{b}\, \overline{(m}^{2})_{32}}{16\pi ^{2}[\overline{m}_{\widetilde{b}_{R}}^{2}-\overline{m}_{\widetilde{s}_{R}}^{2}]}\times \left[ \frac{m_{s}m_{b}}{2M_{w}^{2}\cos ^{2}(\beta )}+\frac{2}{9}\tan ^{2}(\theta _{w})\right] \times \sum ^{1}_{n=0}\frac{(-1)^{n}}{\overline{M}^{2}_{n}}\left[ FC_{2}(X_{\widetilde{\mu }_{n}})\right] .
\end{equation}

The functions \( FC_{1}(X)\, ,\, FC_{2}(X)\, ,\, and\, FN_{1}(X) \) are given
in annex.

The contributions, illustrated in figure1(e-f-g-h), are proportional to \( m_{b} \) due to the left-right diagonal insertion :

\begin{equation}
\label{eq12}
\overline{m}_{LRb}^{2}=-a_{0}m_{0}m_{b}-m_{b}\mu \tan (\beta ).
\end{equation}
 Then from these diagrams, we get:

+for the diagrams 1(e) and 1(f)

\begin{equation}
\label{eq13}
\begin{array}{c}
T_{1e-f}=\frac{g^{2}_{2}g_{3}\, \overline{m}_{LRb}^{2}\, \overline{(m}^{2})_{32}}{16\pi ^{2}[\overline{m}_{\widetilde{b}R}^{2}-\overline{m}_{\widetilde{b}L}^{2}]}\left[ \frac{M_{1}\tan ^{2}(\theta _{w})}{9}+\frac{m_{s}\, m_{b}\, \mu }{2M^{2}_{w}\cos ^{2}(\beta )}\right] \times \sum ^{3}_{n=0}\frac{(-1)^{n}}{\overline{m}^{2}_{n}-\overline{m}^{2}_{\widetilde{sR}}}\times \frac{1}{\overline{M}^{2}_{n1}}\\
\\
\times \left[ \frac{2\times M_{1}^{2}\times FC_{1}(X_{\widetilde{w}_{n1}})}{\overline{M}_{n1}^{2}}+\frac{2\times \mu ^{2}\times FC_{1}(X_{\widetilde{\mu }_{n1}})}{\overline{M}_{n1}^{2}}-3\times FC_{3}(X_{\widetilde{w}_{n1}})-3\times FC_{3}(X_{\widetilde{\mu }_{n1}})\right] 
\end{array}
\end{equation}

where :

\begin{equation}
\label{eq14}
\begin{array}{cc}
\overline{m}_{n}^{2}=\left\{ \begin{array}{c}
\overline{m}_{\widetilde{b}_{R}}^{2}\, for\, n=0,1\\
\overline{m}_{\widetilde{b}_{L}}^{2}\, for\, n=2,3
\end{array}\right.  & \overline{M}_{n1}^{2}=\left\{ \begin{array}{c}
\overline{m}_{\widetilde{b}_{R}}^{2}\, for\, n=0\\
\overline{m}_{\widetilde{s}_{R}}^{2}\, for\, n=1,2\\
\overline{m}_{\widetilde{b}_{L}}^{2}\, for\, n=3
\end{array}\right. 
\end{array}.
\end{equation}

+for the diagrams 1(g) and 1(h) :

\begin{equation}
\label{eq15}
\begin{array}{c}
T_{1g-h}=-\frac{g^{2}_{2}g_{3}\, m_{s}\, m_{b}\, \tan (\theta _{w})\overline{m}_{LRb}^{2}\, \overline{(m}^{2})_{32}}{3\sqrt{2}16\pi ^{2}[\overline{m}_{\widetilde{b}R}^{2}-\overline{m}_{\widetilde{b}L}^{2}]\times [M_{1}^{2}-\mu ^{2}]}\times [\mu +M_{1}\tan (\beta )]^{2}\\
\\
\times \sum ^{3}_{n=0}\frac{(-1)^{n}}{[\overline{m}_{n2}-\overline{m}_{\widetilde{s}R}^{2}]}\times \frac{1}{\overline{M}_{n2}^{2}}[-\frac{\mu ^{2}+2\mu M_{1}}{[M_{1}^{2}-\mu ^{2}]}\{FC_{4}(X_{\widetilde{\mu }_{n2}})-FC_{4}(X_{\widetilde{w}_{n2}})\}\\
\\
+\frac{M^{2}_{1}+2\mu M_{1}}{[M_{1}^{2}-\mu ^{2}]}\{FC_{4}(X_{\widetilde{\mu }_{n2}})-FC_{4}(X_{\widetilde{w}_{n2}})\}+\frac{2(\mu ^{2}+2\mu M_{1})}{\overline{M}_{n2}^{2}}FC_{5}(X_{\widetilde{\mu }_{n2}})-3FC_{6}(X_{\widetilde{\mu }_{n2}})\\
\\
-\frac{2(M^{2}_{1}+2\mu M_{1})}{\overline{M}_{n2}^{2}}FC_{5}(X_{\widetilde{w}_{n2}})+3FC_{6}(X_{\widetilde{w}_{n2}})]
\end{array}
\end{equation}

with:

\begin{equation}
\label{eq16}
\begin{array}{cc}
\overline{m}_{n2}^{2}=\left\{ \begin{array}{c}
\overline{m}_{\widetilde{b}_{L}}^{2}\, for\, n=0,1\\
\overline{m}_{\widetilde{b}_{R}}^{2}\, for\, n=2,3
\end{array}\right.  & \overline{M}_{n2}^{2}=\left\{ \begin{array}{c}
\overline{m}_{\widetilde{b}_{L}}^{2}\, for\, n=0\\
\overline{m}_{\widetilde{s}_{R}}^{2}\, for\, n=1,2\\
\overline{m}_{\widetilde{b}_{R}}^{2}\, for\, n=3
\end{array}\right. .
\end{array}
\end{equation}

The functions \( FC_{3}(X) \),\( FC_{4}(X) \),\( FC_{5}(X) \) and \( FC_{6}(X) \)
are written in annex .

Finally, the contributions drawn on figure 2(a-b) and 2(c-d) are identical with
the last ones (fig.1:e-h), but they are proportional to \( m_{s} \) due to the
left-right diagonal insertion :

\begin{equation}
\label{eq17}
\overline{m}_{LRs}^{2}=-a_{0}m_{0}m_{s}-m_{s}\mu \tan (\beta ).
\end{equation}
 Therefore, the expressions \( T_{2a-b} \) and \( T_{2c-d} \) are obtained
by interchanging : \( \widetilde{b}_{L}\leftrightarrow \widetilde{b}_{R}\, ,\, \widetilde{b}_{R}\leftrightarrow \widetilde{s}_{R}\, ,\, \widetilde{s}_{R}\leftrightarrow \widetilde{s}_{L}\, \, and\, \, m_{b}\leftrightarrow m_{s}\, \, \overline{m}_{LRb}^{2}\leftrightarrow \overline{m}_{LRs}^{2} \)
in the equations(\ref{eq13}, \ref{eq14}, \ref{eq15}, \ref{eq16}) .

\subsection{Gluino contribution.}

Due to gluino-gluon interaction and the multimass insertion method, the eight diagrams (four in the mass insertion method case ) contributing to the chromo-dipole
are illustrated in figure 3 . Due to the flip of chirality on the strange leg, the amplitudes associated to the diagrams(3c)
and (3g) are proportional to \( m_{s} \) . Then, these two contributions are negligible comparatively to the others .

From the diagrams illustrated in figure3 we obtain :

\begin{equation}
\label{eq18}
T_{3a}=\frac{g^{3}_{3}m_{b}(\overline{m}^{2})_{32}}{2N_{c}\times 16\pi ^{2}[\overline{m}_{\widetilde{b}_{R}}^{2}-\overline{m}_{\widetilde{s}_{R}}^{2}]}\times \sum ^{1}_{n=0}\frac{(-1)^{n}}{\overline{M}^{2}_{n}}\left[ FC_{2}(X_{g})\right] 
\end{equation}

with \( X_{g}=\frac{M_{\widetilde{G}}^{2}}{\overline{M}^{2}_{n}}\, ,\, M_{\widetilde{G}} \)
is the gluino mass, \( N_{c}=3 \) and

\begin{equation}
\label{eq19}
\overline{M}_{n}^{2}=\left\{ \begin{array}{c}
\overline{m}_{\widetilde{b}_{R}}^{2}\, for\, n=0\\
\overline{m}_{\widetilde{s}R}^{2}\, for\, n=1,2\\
\overline{m}_{\widetilde{b}_{L}}^{2}\, for\, n=3
\end{array}\right. .
\end{equation}

\begin{equation}
\label{eq20}
\begin{array}{c}
T_{3b}=-\frac{g^{3}_{3}M_{\widetilde{G}}(\overline{m}^{2})_{32}\overline{m}_{LRb}^{2}}{2N_{c}\times 16\pi ^{2}[\overline{m}_{\widetilde{b}R}^{2}-\overline{m}_{\widetilde{b}L}^{2}]}\times \sum ^{3}_{n=0}\frac{(-1)^{n}}{\overline{m}^{2}_{n}-\overline{m}_{\widetilde{s}R}^{2}}\\
\times \frac{1}{\overline{M}_{n}^{2}}\left[ \frac{2M^{2}_{\widetilde{G}}}{\overline{M}_{n}^{2}}FC_{1}(X_{g})-3FC_{3}(X_{g})\right] 
\end{array}
\end{equation}

and

\begin{equation}
\label{eq21}
\begin{array}{c}
T_{3d}=-\frac{g^{3}_{3}M_{\widetilde{G}}(\overline{m}^{2})_{32}\overline{m}_{LRs}^{2}}{2N_{c}\times 16\pi ^{2}[\overline{m}_{\widetilde{s}R}^{2}-\overline{m}_{\widetilde{b}R}^{2}]}\times \sum ^{3}_{n=0}\frac{(-1)^{n}}{\overline{m}^{2}_{n_{3}}-\overline{m}_{\widetilde{s}L}^{2}}\\
\times \frac{1}{\overline{M}_{n3}^{2}}\left[ \frac{2\times M_{\widetilde{G}}^{2}}{\overline{M}_{n3}^{2}}FC_{1}(X_{g3})-3FC_{3}(X_{g3})\right] 
\end{array}
\end{equation}

with

\begin{equation}
\label{eq22}
\begin{array}{cc}
\overline{m}_{n3}^{2}=\left\{ \begin{array}{c}
\overline{m}_{\widetilde{s}R}^{2}\, for\, n=0,1\\
\overline{m}_{\widetilde{b}R}^{2}\, for\, n=2,3
\end{array}\right.  & \overline{M}_{n3}^{2}=\left\{ \begin{array}{c}
\overline{m}_{\widetilde{s}R}^{2}\, for\, n=0\\
\overline{m}_{\widetilde{s}L}^{2}\, for\, n=1,2\\
\overline{m}_{\widetilde{b}R}^{2}\, for\, n=3
\end{array}\right. .
\end{array}
\end{equation}

From the others diagrams we deduce :

\begin{equation}
\label{eq23}
T_{3e}=-\frac{N_{c\, }g^{3}_{3}\, m_{b}\, (\overline{m}^{2})_{32}}{2\times 16\pi ^{2}[\overline{m}_{\widetilde{b}R}^{2}-\overline{m}_{\widetilde{s}R}^{2}]}\times \sum ^{1}_{n=0}\frac{(-1)^{n}}{\overline{M}^{2}_{n}}\left[ FC_{2}(X_{g})\right] 
\end{equation}

\begin{equation}
\label{eq24}
\begin{array}{c}
T_{3f}=-\frac{N_{c\, }g^{3}_{3}\, M_{\widetilde{G}}\, (\overline{m}^{2})_{32}\overline{\, m}_{LRb}^{2}}{2\times 16\pi ^{2}[\overline{m}_{\widetilde{b}R}^{2}-\overline{m}_{\widetilde{b}L}^{2}]}\times \sum ^{3}_{n=0}\frac{(-1)^{n}}{\overline{m}_{n2}^{2}-\overline{m}_{\widetilde{s}R}^{2}}\\
\times \frac{1}{\overline{M}_{n2}^{2}}\left[ \frac{2}{\overline{M}_{n2}^{2}}G_{1}(X_{g2})+3FC_{3}(X_{g2})\right] 
\end{array}
\end{equation}

\begin{equation}
\label{eq25}
\begin{array}{c}
T_{3h}=-\frac{N_{c}\, g^{3}_{3}\, M_{\widetilde{G}}\, (\overline{m}^{2})_{32}\, \overline{m}_{LRs}^{2}}{2\times 16\pi ^{2}[\overline{m}_{\widetilde{s}R}^{2}-\overline{m}_{\widetilde{b}R}^{2}]}\times \sum ^{3}_{n=0}\frac{(-1)^{n}}{\overline{m}^{2}_{n_{4}}-\overline{m}_{\widetilde{s}L}^{2}}\\
\times \frac{1}{\overline{M}_{n4}^{2}}\left[ \frac{2}{\overline{M}_{n4}^{2}}G_{1}(X_{g4})+3G2(X_{g4})\right] 
\end{array}
\end{equation}

with 
\begin{equation}
\label{eq26}
\begin{array}{cc}
\overline{m}_{n4}^{2}=\left\{ \begin{array}{c}
\overline{m}_{\widetilde{b}R}^{2}\, for\, n=0,1\\
\overline{m}_{\widetilde{s}R}^{2}\, for\, n=2,3
\end{array}\right.  & \overline{M}_{n4}^{2}=\left\{ \begin{array}{c}
\overline{m}_{\widetilde{b}R}^{2}\, for\, n=0\\
\overline{m}_{\widetilde{s}L}^{2}\, for\, n=1,2\\
\overline{m}_{\widetilde{s}R}^{2}\, for\, n=3
\end{array}\right. .
\end{array}
\end{equation}

Due to the multimass insertion method, all the amplitudes depend on the squark left and right masses (contrary to the mass insertion 
method case where an average squark mass contribute ) . Now, we can build the SUSY contribution to the decay amplitude : \( M_{B_{d}\rightarrow \phi K^{0}}^{(SUSY)} \).

\section{Numerical results on SUSY contribution to the decay phase}

In the decay amplitude, we take into account the chromo-dipole term only . Then,
with the factorization approximation\cite{11}-\cite{12}, the decay amplitude(\ref{eq9})
takes the following form :

\begin{equation}
\label{eq27}
M_{B_{d}\rightarrow \phi K^{0}}=2\times (P_{B}\epsilon _{\phi })m_{\phi }^{2}f_{\phi }F_{+}^{BK}\frac{1}{16}(1-\frac{1}{N_{c}^{2}})\left[ g_{3}K_{DM}C_{R}^{DM}+(L\leftrightarrow R)\right] 
\end{equation}

\begin{equation}
\label{eq28}
\begin{array}{c}
where\: <\Phi (P_{\phi },\varepsilon _{\phi })|\overline{s}^{a}\gamma ^{\mu }s^{a}|0>=m_{\phi }^{2}f_{\phi }\varepsilon ^{\mu }\\
and\: <K^{0}(P_{K})|\overline{s}^{a}\gamma ^{\mu }b^{a}|\overline{B}_{d}(P_{B})>=F_{+}^{BK}(P_{B}+P_{K})^{\mu }+F_{-}^{BK}(P_{B}-P_{K})^{\mu }
\end{array}.
\end{equation}

The \( K_{DM} \) coefficient is induced from the hadronization of the chromo-dipole
moment . From the quark model and the heavy quark effective theory, we have
:

\begin{equation}
\label{eq29}
K_{DM}\approx \frac{m_{b}^{2}}{2q^{2}}\left[ \frac{9}{8}+O(m_{\phi }^{2}/m_{b}^{2})\right] ,
\end{equation}
 where q is the momentum transfer in the gluon line such that :

\begin{equation}
\label{eq30}
q^{2}=\frac{1}{2}(m_{B}^{2}-\frac{1}{2}m_{\phi }^{2}+m_{K}^{2}),
\end{equation}

and \( K_{DM}\approx 1.2\: for\: \alpha _{s}(m_{Z})=0.118\: ,\: \alpha _{em}(m_{Z})=1/128\: and\: m_{b}=5Gev \)
. In the decay process \( B_{d}\longrightarrow \Phi K_{s} \) the CP violation
is determined by :

\begin{equation}
\label{eq31}
\phi ^{total}=\phi _{B_{d}\rightarrow \overline{B}_{d}}^{mix}+arg\left[ M_{\overline{B}_{d}\rightarrow \phi K^{0}}^{SM}\: +M_{\overline{B}_{d}\rightarrow \phi K^{0}}^{(SUSY)}\right] .
\end{equation}

So, the SUSY contribution to the decay phase is expressed by :

\begin{equation}
\label{eq32}
\tan (\Delta \phi _{\overline{B}_{d}\rightarrow \phi K^{0}}^{decay})\equiv \frac{\left| M_{\overline{B}_{d}\rightarrow \phi K^{0}}^{SUSY}\right| }{\left| M_{\overline{B}_{s}\rightarrow \phi K^{0}}^{SM}\right| },
\end{equation}

where \( \Delta \phi _{\overline{B}_{d}\rightarrow \phi K^{0}}^{decay} \) is
the maximal possible correction to the decay phase such that:

\begin{equation}
\label{eq33}
\phi ^{(L)}_{3}-\phi ^{(L)}_{2}\simeq arg\left[ M_{\overline{B}_{d}\rightarrow \phi K^{0}}^{SM}\right] +\frac{\pi }{2}.
\end{equation}

The standard model contribution \cite{12} has the following form :

\begin{equation}
\label{eq34}
\begin{array}{c}
M^{SM}=-\frac{\pi \alpha _{w}}{2M_{w}^{2}}V_{tb}V_{ts}^{*}m_{\phi }^{2}f_{\phi }\varepsilon ^{\mu }\times 2P^{\mu }_{B}F^{BK}_{+}\times \\
\left[ C+\frac{\alpha _{s\, }(N^{2}_{c}-1)\, c_{11\, }m_{b}}{16\pi ^{2}\, K^{2}\, N^{2}_{c}}\left( m_{b}+4(1+\frac{m_{s}^{2}}{m_{\phi }^{2}})-m_{s}(1-2\frac{m^{2}_{b}+2m_{\phi }^{2}-m_{s}^{2}}{m_{\phi }^{2}})-\frac{4m_{s}m_{b}}{m_{\phi }^{2}}\right) \right] 
\end{array}
\end{equation}
 \vspace{0.5cm}

with \( C=c_{3}+c_{4}+c_{5}+\frac{1}{N_{c}}(c_{3}+c_{4}+c_{6})-\frac{1}{2}\left[ c_{7}+c_{9}+c_{10}+\frac{1}{N_{c}}(c_{8}+c_{9}+c_{10})\right]  \)
and the \( c_{i} \) are the Wilson coefficients . The equation (\ref{eq32})
is very sensitive to the standard model contribution . So we choose to take
into account the next leading order QCD corrections, with the renormalization
scale :\( \mu =m_{b}=5Gev\: , \: \alpha _{em}(m_{Z})=1/128\: , \: \alpha _{s}(m_{Z})=0.118\:  and  \: m_{t}=176Gev \)
. So, the Wilson coefficients have the following values : \vspace{0.5cm}

\( \begin{array}{cccc}
c_{3}=0.0174 & c_{4}=-0.0373 & c_{5}=0.0104 & c_{6}=-0.0459\\
c_{7}=1.398\times 10^{-5} & c_{8}=3.919\times 10^{-4} & c_{9}=-0.0103 & c_{10}=1.987\times 10^{-3},
\end{array} \) \vspace{0.5cm}

and the DPO coefficient at the two loop level is \( c_{11}=-0.299. \)

Concerning the SUSY contributions, we have considered:

+the neutralino contribution

\begin{equation}
\label{eq35}
M_{\overline{B}_{d}\rightarrow \phi K^{0}}^{\chi }=2(P_{B}\varepsilon _{\phi })m_{\phi }^{2}f_{\phi }F_{+}^{BK}\times \frac{1}{8}(1-\frac{1}{N_{c}^{2}})g_{3}K_{DM}\times \frac{1}{m_{b}}\times T_{\chi }
\end{equation}

with \( T_{\chi }=T_{1a-b}+T_{1c-d}+T_{1e-f}+T_{1g-h}+T_{2a-b}+T_{2c-d} \)
and

\begin{equation}
\label{eq36}
\tan (\Delta \phi _{\overline{B}_{d}\rightarrow \phi K^{0}}^{decay})_{\chi }\equiv \frac{\left| M_{\overline{B}_{d}\rightarrow \phi K^{0}}^{\chi }\right| }{\left| M_{\overline{B}_{s}\rightarrow \phi K^{0}}^{SM}\right| }.
\end{equation}

+the gluino contribution

\begin{equation}
\label{eq37}
M_{\overline{B}_{d}\rightarrow \phi K^{0}}^{\widetilde{G}}=2(P_{B}\varepsilon _{\phi })m_{\phi }^{2}f_{\phi }F_{+}^{BK}\times \frac{1}{8}(1-\frac{1}{N_{c}^{2}})g_{3}K_{DM}\times \frac{1}{m_{b}}\times T_{\widetilde{G}}
\end{equation}

with \( T_{\widetilde{G}}=T_{3a}+T_{3b}+T_{3d}+T_{3e}+T_{3f}+T_{3h} \) and

\begin{equation}
\label{eq37}
\tan (\Delta \phi _{\overline{B}_{d}\rightarrow \phi K^{0}}^{decay})_{\widetilde{G}}\equiv \frac{\left| M_{\overline{B}_{d}\rightarrow \phi K^{0}}^{\widetilde{G}}\right| }{\left| M_{\overline{B}_{s}\rightarrow \phi K^{0}}^{SM}\right| }.
\end{equation}

For each case and different \( K_{DM} \) values, we have calculated the decay
phase . From the atmospheric result, we take the tau neutrino mass : \( m_{\nu _{\tau }}=0.07eV \)
and \( V_{L32}=-\frac{1}{\sqrt{2}} \) . The right-handed neutrino mass is constrained
by the neutrino Yukawa coupling eq(\ref{eq4}) and the \( Br\left( \mu \rightarrow e\gamma \right)  \)
experimental bound\cite{6} . So we choose \( M_{\nu _{R}}=2\times 10^{14}GeV \)
. In our numerical analysis we have used the suspect2\footnote{%
kneur@lpm.univ-montp2.fr 
} program by which is possible to obtain the sparticle mass spectrum for different
values on SUSY parameters that is :\( m_{0}\: ,\: m_{1/2}\: ,\: \tan \beta \: ,\: sign(\mu ) \)
. In this program RG evolution of parameters is taken into account . To fix the gluino mass we take $m_{1/2}= 200 \mbox{ Gev}$ and 
$ sign(\mu)$ positive . These values are given in tables:\ref{tab1}, \ref{tab2}, \ref{tab3} and are consistent with 
\( Br\left( \mu \rightarrow e\gamma \right)  \) experimental bound . 
\begin{table}
\centering
\begin{tabular}{|c|c|c|c|c|c|c|c|c|c|c|}
\hline 
\( m_{0} \)&
 200&
 300&
 400&
 500&
 600&
 700&
 800&
 1000&
 1200&
 1500\\
\hline 
\( M_{1} \)&
 81.55&
 81.79&
 82&
 82.27&
 82.56&
 82.81&
 83&
 83.48&
 83.9&
 84.3\\
\hline 
\( \mu  \)&
 309.1&
 327.7&
 352.5&
 381.5&
 413.7&
 448.2&
 483.5&
 564.5&
 640.9&
 761.1\\
\hline 
\( m_{\widetilde{s}_{R}} \)&
 466.7&
 515.5&
 577.6&
 648.8&
 727.5&
 810.5&
 897.2&
 1078&
 1264&
 1550\\
 \( m_{\widetilde{s}_{L}} \)&
 484.6&
 531.7&
 592&
 661.7&
 738.9&
 820.8&
 906.5&
 885.6&
 1271&
 1556\\
\hline 
\( m_{\widetilde{b}_{R}} \)&
 466.4&
 515.1&
 577.2&
 648.3&
 726.9&
 809.9&
 896.5&
 1077&
 1263&
 1549\\
 \( m_{\widetilde{b}_{L}} \)&
 433.6&
 466.6&
 510&
 561.1&
 619.1&
 681.3&
 747.4&
 885.6&
 1032&
 1259\\
\hline 
\( M_{\widetilde{G}} \)&
 511.7&
 518.2&
 525.1&
 531.2&
 538.4&
 544.6&
 550.6&
 561&
 570.6&
 582.6  \\
\hline 
\end{tabular}
\par{}

\caption{\label{tab1}Sparticle mass spectrum (en GeV) for \protect\protect\protect\( \mu >0\protect \protect \protect \),\protect\protect\protect\( m_{1/2}=200GeV\protect \protect \protect \)
and \protect\protect\protect\( \tan \beta =3\protect \protect \protect \)}
\end{table}
\begin{table}
\centering 

\begin{tabular}{|c|c|c|c|c|c|c|c|c|c|c|}
\hline 
\( m_{0} \)&
 200&
 300&
 400&
 500&
 600&
 700&
 800&
 1000&
 1200&
 1500\\
\hline 
\( M_{1} \)&
 81.48&
 81.75&
 82.06&
 82.32&
 82.48&
 82.74&
 82.96&
 83.4&
 83.75&
 84.31\\
\hline 
\( \mu  \)&
 263.7&
 265.6&
 267.7&
 270.3&
 272&
 274.4&
 277.3&
 284.9&
 278.2&
 265.5\\
\hline 
\( m_{\widetilde{s}_{R}} \)&
 465.8&
 514.9&
 577.1&
 648.7&
 726.7&
 809.9&
 896.5&
 1077&
 1264&
 1550\\
 \( m_{\widetilde{s}_{L}} \)&
 483.7&
 531.1&
 591.6&
 661.6&
 738.1&
 820.2&
 905.8&
 1085&
 1270&
 1555\\
\hline 
\( m_{\widetilde{b}_{R}} \)&
 462.9&
 511.5&
 573&
 643.8&
 721&
 803.4&
 889.2&
 1068&
 1253&
 1537\\
 \( m_{\widetilde{b}_{L}} \)&
 435&
 469.5&
 514.4&
 567.2&
 626.1&
 690&
 757.3&
 899.1&
 1049&
 1280\\
\hline 
\( M_{\widetilde{G}} \)&
 511.3&
 518.3&
 525.3&
 532.2&
 538.1&
 544.4&
 550&
 560.8&
 570.2&
 583.3  \\
\hline 
\end{tabular}
\par{}

\caption{\label{tab2}Sparticle mass spectrum (en GeV) for \protect\protect\protect\( \mu >0\protect \protect \protect \),\protect\protect\protect\( m_{1/2}=200GeV\protect \protect \protect \)
and \protect\protect\protect\( \tan \beta =10\protect \protect \protect \)}
\end{table}
\begin{table}
\centering 

\begin{tabular}{|c|c|c|c|c|c|c|c|c|c|c|}
\hline 
\( m_{0} \)&
 200&
 300&
 400&
 500&
 600&
 700&
 800&
 1000&
 1200&
 1500\\
\hline 
\( M_{1} \)&
 81.53&
 81.8&
 82.02&
 82.27&
 82.55&
 82.81&
 82.99&
 83.5&
 83.81&
 84.24\\
\hline 
\( \mu  \)&
 258.8&
 258.8&
 258.7&
 258.7&
 257.8&
 256.2&
 255.1&
 253.4&
 237.7&
 190.8\\
\hline 
\( m_{\widetilde{s}_{R}} \)&
 466&
 515.2&
 576.9&
 648.5&
 727&
 810.1&
 896.7&
 1077&
 1264&
 1550\\
 \( m_{\widetilde{s}_{L}} \)&
 483.9&
 531.4&
 591.4&
 661.3&
 738.5&
 820.4&
 906&
 1085&
 1270&
 1555\\
\hline 
\( m_{\widetilde{b}_{R}} \)&
 444.1&
 488.4&
 544.3&
 609.2&
 680.7&
 756.3&
 835.5&
 1001&
 1172&
 1436\\
 \( m_{\widetilde{b}_{L}} \)&
 425.8&
 457.8&
 499.2&
 548.6&
 604.3&
 664&
 727.3&
 861&
 1002&
 1222\\
\hline 
\( M_{\widetilde{G}} \)&
 511&
 517.9&
 524.2&
 531&
 537.9&
 544&
 549.6&
 561&
 569.8&
 582.1  \\
\hline 
\end{tabular}
\par{}

\caption{\label{tab3}Sparticle mass spectrum (en GeV) for \protect\protect\protect\( \mu >0\protect \protect \protect \),\protect\protect\protect\( m_{1/2}=200GeV\protect \protect \protect \)
and \protect\protect\protect\( \tan \beta =30\protect \protect \protect \)}
\end{table}

We plot in figures 4-6, the neutralino contribution to \( \tan (\Delta \phi _{\overline{B}_{d}\rightarrow \phi K^{0}}^{decay})_{\chi }\equiv \frac{\left| M_{\overline{B}_{d}\rightarrow \phi K^{0}}^{\chi }\right| }{\left| M_{\overline{B}_{s}\rightarrow \phi K^{0}}^{SM}\right| } \)
and gluino ones eq(\ref{eq37}) on figures 7-9 .\\
From the results obtained, we remark the following .\\

(1) Because of our use of multimass insertion method, the results depend on the gluino, wino, higgsino and squark left and right masses 
(tables \ref{tab1}, \ref{tab2}, \ref{tab3} ) .\\
(2) The gluino and neutralino contributions have the same behaviour and the upper values are obtained for $m_{0}= 500 \mbox{ Gev}$ 
(figures 4-9) .\\
(3) As is shown on fig.4 and fig.7, these two contributions are very sensitive to tan$\beta$ and $K_{DM}$ values . 
They increase with the enhancement of these two parameters .\\
(4) The dependency on $a_{0}$ of the decay phase is presented on fig.6 and fig.9 . The phase becomes more important with increasing 
$a_{0}$ parameter .\\
(5) In the gluino case, thanks to the gluino mass (500 GeV), the decay phase is significant for large tan$\beta$ and moderate $K_{DM}$ values .
 Indeed, from fig.8, for $a_{0}= 0$, tan$\beta= 30$ and $K_{DM}= 1.5$ the upper value of the decay phase is $4.10^{-2}$ . 
This value is greater than the uncertainties in the SM calculation \( O(10^{-2}) \) \cite{8} . Moreover, as is shown on fig.9, 
for $a_{0}= 10$ the gluino contribution is more competitive with the SM one because the upper value is 0.1 . Our results are consistent with
those of the reference\cite{9} (figure 2) . However, our results are much smaller due to multimass insertion method and the value of the 
\( M_{\nu _{R}}=2\times 10^{14}GeV \) (instead of \( M_{\nu _{R}}=5\times 10^{14}GeV \) in the reference\cite{9}) .\\
(6) For the neutralino case, due to the small values of $\mu$ and $M_{1}$ (tables \ref{tab1}, \ref{tab2}, \ref{tab3} ), the decay phase 
(fig.4,5) cannot be competitive with the gluino one for $a_{0}$ less than 250 (even for large tan$\beta$ and $K_{DM}$ values) . 
The highest value ($10^{-3}$) is very small in comparison with the uncertainties in the SM model calculation . Nevertheless, for large $a_{0}$ 
(\( a_{0}\geq 250 \)) and tan$\beta$ values, we obtain results (fig.6) greater than $10^{-2}$ . Then, the neutralino contribution changes the phase in 
the process \( B_{d}\longrightarrow \Phi K_{s} \) .

.
.

\section{Conclusion}

In the context of \( SU(5) \) grand unified theory with right-handed neutrino,
we have calculated the neutralino and gluino penguin contributions to the phase
in \( B_{d}\longrightarrow \Phi K_{s} \) . In our calculation, we apply the multimass insertion method . 
For convenient choices of SUSY parameters,
the neutralino contribution to the decay phase is greater than \( O(10^{-2}) \)
. For the gluino contribution the decay phase is the order of 0.1 . Then, the
SUSY contributions change the decay phase in the process \( B_{d}\longrightarrow \Phi K_{s} \)
.

\section{Annex}

Now, we give analytic expressions of the Feynman integrals which appear in the
evaluation of the amplitudes :

\( \begin{array}{c}
\\
G_{1}(X)=\frac{-4-11X-14X\log (X)+16X^{2}-4X^{2}\log (X)-X^{3}}{2X(X-1)^{4}}\\

\end{array} \),

\( \begin{array}{c}
\\
G_{2}(X)=\frac{-38-24\log (X)+81X+18X\log (X)-54X^{2}+11X^{3}}{36(X-1)^{4}}\\

\end{array} \),

\( \begin{array}{c}
\\
FC_{5}(X)=\frac{1+9X+6X\log (X)-9X^{2}+6X^{2}\log (X)-X^{3}}{3(X-1)^{5}}\\

\end{array} \),

\( \begin{array}{c}
\\
FC_{4}(X)=\frac{-1+6X-3X^{2}+6X^{2}\log (X)-X^{3}}{6(X-1)^{4}}\\

\end{array} \),

\( \begin{array}{c}
\\
\begin{array}{c}
FC_{1}(X)=\frac{5+2\log (X)-4X+4X\log (X)-X^{2}}{2(X-1)^{4}}\\
\\
FC_{2}(X)=\frac{-1+-X-3X^{2}+6X^{2}\log (X)-2X^{3}}{6(X-1)^{4}}\\
\\
FC_{3}(X)=\frac{-2-3X-6X\log (X)+6X^{2}-X^{3}}{6(X-1)^{4}}\\

\end{array}\\

\end{array} \)

\( \begin{array}{c}
\\
FC_{6}=\frac{1-8X-12X^{2}\log (X)+8X^{3}-X^{4}}{12(X-1)^{5}}\\

\end{array} \),

\( \begin{array}{c}
\\
FN_{1}=\frac{1+2X\log (X)-X^{2}}{2(X-1)^{3}}\\

\end{array} \).

\vspace{0.5cm}

ACKNOWLEDGMENTS: I would like to thank A.Djouadi, S.Lavignac and G.Mennessier
for helpful comments . This work was supported in part by Euro-Gdr SUSY.

\end{document}